\documentstyle[prb,preprint,aps]{revtex}
\begin{document}
\draft

\title{Soluble `Supersymmetric' Quantum XY Model}

\author{A.E. Rana and S.M. Girvin}

\address{Department of Physics, Indiana University, Bloomington\ \ IN\ \
47405}

\date{\today}

\maketitle
\begin{abstract}
We present a `supersymmetric' modification of the $d$-dimensional quantum
rotor
model whose ground state is exactly soluble.  The model undergoes a
vortex-binding transition from insulator to metal as the rotor coupling is
varied.  The Hamiltonian contains three-site terms which are relevant: they
change the universality class of the transition from that of the ($d+1$)---
to
the $d$-dimensional classical XY model.  The metallic phase has algebraic
ODLRO
but the superfluid density is identically zero.  Variational wave functions
for
single-particle and collective excitations are presented.
\end{abstract}


\narrowtext
This paper discusses an exactly soluble modified version of the quantum XY
model
\begin{equation}
{\cal H}_0=-U\sum_j \frac{\partial^2}{\partial\theta_j^2}-J\sum_{j\delta}
\cos{(\theta_j-\theta_{j+\delta})},
\label{one}
\end{equation}
where $j$ is summed on sites of a hypercubic lattice and $\delta$ is summed
on
near-neighbor vectors.\cite{referencea}  This model is often used to
describe
the superconductor-insulator transition in granular superconductors and
Josephson junction arrays.\cite{SIT}  The coupling constant $J$ represents
the
strength of the Josephson coupling between the order parameter phases
$\theta_i$
and $\theta_j$ on neighboring grains.  The parameter $U$ represents the
charging
energy of the grains.  The boson (Cooper pair) number operator conjugate to
the
phase is the angular momentum ${\hat n}_j\equiv -i\partial_{\theta_j}$.
While
this is correctly quantized in integer values, it can be negative.  Thus
the
model implicitly assumes a large background number $n_0$ of bosons per
lattice
site so that ${\hat n}_j$ represents local deviations (positive or
negative)
from this mean (integer) value.  We can view the cosine term as a mutual
torque
which transfers quanta of angular momentum (bosons) from one site to the
next.
Thus the quantum XY model is essentially equivalent to (i.e., in the same
universality class as) the boson Hubbard model.  For large $U/J$ the ground
state is a Mott-Hubbard insulator and for small $U/J$ it is superfluid
which
exhibits off-diagonal long range order (ODLRO) in the phase field
correlations
(at zero temperature)
\begin{equation}
G_{ij}\equiv\langle e^{i\theta_i}e^{-i\theta_j}\rangle
\label{two}
\end{equation}
for dimension $d>1$ (and algebraic ODLRO for $d=1$).  The transition
between
the superfluid and Mott-Hubbard insulating states is continuous and is in
the
universality class of the $d+1$-dimensional {\em classical\/} XY
model.\cite{SIT,donaich,fisherfisher}  The extra dimension arises from the
fact that in the path integral representation of the partition function,
the
Euclidean time interval $0\leq\tau\leq\hbar\beta$ diverges at zero
temperature.

While the physics of this model is now completely understood, it has
resisted
exact solution in all dimensions.  It is interesting to consider a
Jastrow-like
variational wave function
\begin{equation}
\psi_0(\theta_1,\cdots ,\theta_N)=\exp{\Biggl[
-\Bigl(\frac{\lambda}{U}\Bigr)
V(\theta_1,\cdots ,\theta_N)\Biggr]}
\label{three}
\end{equation}
where $\lambda$ is a variational parameter and
\begin{equation}
V\equiv -J\sum_{j\delta} \cos{(\theta_j-\theta_{j+\delta})}
\label{four}
\end{equation}
is the potential energy from eq.(\ref{one}).  This form is motivated by the
harmonic oscillator
\begin{equation}
H=\frac{p^2}{2m}+\frac{1}{2}Kx^2
\label{five}
\end{equation}
for which the exact ground state is
\begin{equation}
\psi (x)=\exp{\Biggl[ -\hbar\omega \Bigl(\frac{1}{2}Kx^2\Bigr)\Biggr]}.
\label{six}
\end{equation}
The variational state in eq.(\ref{three}) is thus in the spirit of the
harmonic
spin-wave approximation in which one expands the cosine term to second
order in
deviations from the classical ground state.  The wave function is much better
than this however because it obeys the correct periodicity under
$\theta_j\rightarrow\theta_j+2\pi$.  This feature is crucial to the
existence
of (quantum) vortices in the ground state and hence allows for the
possibility
of a phase transition to the insulating state--physics which is completely
missing from the spin-wave approximation.

The purpose of the present paper is to examine the variational wave
function of
eq.(\ref{three}) and to consider a Hamiltonian for which $\psi_0$ happens
to be
the exact ground state.\cite{arovas}  The question of what Hamiltonians
have
Jastrow wave functions for ground states has a long
history\cite{feenburg,sutherland} and this question has recently been
reexamined from a modern perspective by Kane, et al.\cite{kaneetal}  Except
for
special cases,\cite{sutherland} the generic requirement is that the
Hamiltonian have three-body interactions of a particular form.  We will see
shortly that the analog of this for the present problem is three-site
interactions.  Kane, et al. argue that these three-body interactions are
(perturbatively) irrelevant in the renormalization group sense.  That is,
they
have no effect on the long-distance properties of the system (other than a
trivial renormalization of the speed of sound).  This result is
perturbative because it neglects vortex-like excitations.\cite{kaneetal}
While, strictly speaking, the above statements are true, they can be quite
misleading if the ground state of the system undergoes a phase transition.
We
show explicitly below that the universality class of the transition is
completely altered by the inclusion of particular three-site interactions.
This
is a non-perturbative effect precisely due to the role played by vortices
in the
(zero temperature) transition.  The discussion below in readily generalized
to
any dimension, but for definiteness we consider only the case $d=2$.

We construct the desired Hamiltonian by defining the operators
\begin{mathletters}
\begin{eqnarray}
Q_j&=&\sqrt{U}{\partial\over\partial\theta_j}+{J\over\sqrt{U}}\sum_\delta
\bigl[\sin{(\theta_j-\theta_{j+\delta})} + \sin{(\theta_j -
\theta_{j-\delta})}\bigr] \label{sevena}\\
Q_j^\dagger
&=&-\sqrt{U}{\partial\over\partial\theta_j}+{J\over\sqrt{U}}\sum_\delta
\bigl[\sin{(\theta_j-\theta_{j+\delta})} + \sin{(\theta_j -
\theta_{j-\delta})}\bigr].\label{sevenb}
\end{eqnarray}
\end{mathletters}
It is readily verified using Eq.~(\ref{three}) that
\begin{equation}
Q_j\psi_0=0
\label{eight}
\end{equation}
for every $j$ (we take $\lambda =1$ hereafter).

The `supersymmetric'\cite{solubleanyon,arovas} Hamiltonian
\begin{equation}
H=\sum_jQ_j^\dagger Q_j^{\phantom{\dagger}}
\label{nine}
\end{equation}
is clearly positive semidefinite and therefore $\psi_0$ is an exact,
zero-energy ground state of $H$.  Using eq.(\ref{sevena}) we can write $H$
in the form
\begin{eqnarray}
H&=&-U\sum_j \partial_{\theta_j}^2-J\sum_{j\delta}
\cos{(\theta_j-\theta_{j+\delta})}\nonumber\\
&&+{J^2\over U}\sum_{j\delta\delta
'}\sin{(\theta_j-\theta_{j+\delta})}\sin{(\theta_j-\theta_{j+\delta
'})}.\label{ten}
\end{eqnarray}
The first two terms on the righthand side are equivalent to the usual
quantum XY
model of eq.(\ref{one}).  The remaining term is a perturbation consisting
of
two- and three-site interactions.  These terms represent the simultaneous
hopping of a pair of bosons.  The $\delta =\delta'$ terms give rise to a
$\cos{2\theta}$ coupling, a form which has been studied by Lee and
Grinstein.\cite{highercharge}

Now that we have the Hamiltonian and the exact ground state, let us examine
the
nature of the ground state as a function of the quantum fluctuation
parameter
$U/J$ to see if the system undergoes a phase transition.  The
(unnormalized)
probability distribution of the phase angles is
\begin{equation}
P[\theta_1,\theta_2,\cdots,
\theta_N]\equiv\vert\psi_0\vert^2=\exp{\Biggl\lbrace\frac{2J}{U}\sum_{j\delta}
\cos{(\theta_j-\theta_{j+\delta})}\Biggr\rbrace }
\label{eleven}
\end{equation}
which is identical to the Boltzmann factor for the {\em classical\/} 2D XY
model
with $U/2J$ playing the role of dimensionless temperature.  This model
undergoes
a Kosterlitz-Thouless phase transition at a critical
temperature\cite{olsson}
$U/2J\approx 0.9$.  This is clearly a different universality class from
that of
the usual 2D quantum XY model which is known to be in the universality
class of
the 3D XY model.  Thus we see that the three site terms are relevant to the
transition (at least when they have the particular strength given in
eq.(\ref{ten})).

If we knew the exact ground state $\Phi (\theta_1,\cdots ,\theta_N)$ of the
usual quantum rotor problem, then $\vert\Phi\vert^2$ would of course define
a 2D
classical statistical mechanics problem.  However, the fake classical
Hamiltonian would necessarily contain long-range forces (in order to give
the
3D XY universality class in a 2D model\cite{arovas}).

We are used to the notion that thermal fluctuations produce vortices.  Here
we
see a nice illustration of the fact that even at zero temperature, vortices
can
be produced by quantum fluctuations.  For $U/2J>T_{\rm KT}^*$ the largest
amplitude configurations in the ground state contain free vortices and the
spin-spin correlation function decays exponentially
\begin{equation}
G_{ij}\equiv\langle e^{i\theta_i}e^{-i\theta_j}\rangle\sim e^{-\vert{\vec
R}_i-{\vec R}_j\vert /\xi},
\label{twelve}
\end{equation}
whereas for $U/2J<T_{\rm KT}^*$ the correlations decay only algebraically
\begin{equation}
G_{ij}\sim\vert {\vec R}_i-{\vec R}_j\vert^{-\eta}
\label{thirteen}
\end{equation}
because vortices are confined.  That is, virtual vortex-antivortex `vacuum
fluctuations' appear but do not proliferate.

Knowing the ground state exactly, we now find approximate excited states of
the Hamiltonian. Let us assume an excited state loosely analogous to the
Feynman-Bijl form\cite{feynman}
\begin{equation}
\Psi_{\vec{q}}=\rho_{\vec{q}}\Psi_0.
\label{fourteen}
\end{equation}
As discussed below, the form of $\rho_{\vec{q}}$ depends on the choice of
excitation. If the excited state is orthogonal to the ground state, i.e.,
\begin{equation}
\langle\Psi_{\vec{q}}\vert\Psi_0\rangle =0
\label{fifteen}
\end{equation}
then we can use the variational principle to find an upper bound on the
excitation  energy (taking advantage of the fact that the ground state
energy
vanishes)
\begin{equation}
\Delta_{\vec{q}}\leq\frac{\langle\Psi_{\vec{q}}\vert
H\vert\Psi_{\vec{q}}\rangle}{\langle\Psi_{\vec{q}}\vert\Psi_{\vec{q}}\rangle
}
\label{sixteen}
\end{equation}
where $\langle\Psi_{\vec{q}}\vert H\vert\Psi_{\vec{q}}\rangle =f(\vec{q})$
is the
`oscillator strength', and $\langle\Psi_q\vert\Psi_q\rangle =s(\vec{q})$ is
the
`static structure factor'. Writing $\Psi_{\vec{q}}$ is terms of $\Psi_0$:
\begin{eqnarray}
f(\vec{q})&=&\langle\Psi_{\vec{q}}\vert H\vert\Psi_{\vec
q}\rangle\nonumber\\
&=&\langle\Psi_0\vert \rho_{-\vec{q}}H\rho_{\vec
q}\vert\Psi_0\rangle ,\label{seventeen}
\end{eqnarray}
and writing the Hamiltonian in terms of $Q$'s and using the fact that
$Q_j\Psi_0=0$,
\begin{equation} f({\vec
q})=\langle\Psi_0\vert [\rho_{-\vec{q}},Q_j^\dagger] [Q_j,\rho_{\vec{q}}]
\vert\Psi_0\rangle
\label{eighteen}
\end{equation}

By substituting an explicit form of $\rho_{\vec{q}}$ in the above results,
$f(\vec{q})$ can be calculated immediately.

First we consider an excited state wave function which describes a
single-particle excitation:
\begin{equation}
\Psi_{\vec{q}}=b_q^\dagger\Psi_0
\label{nineteen}
\end{equation}
where:
\begin{equation}
b_{\vec{q}}^\dagger ={1\over\sqrt{N}}\sum_j e^{i\vec{q}\cdot {\vec
R}_j}e^{i\theta_j}
\label{twenty}
\end{equation}
is the spatial Fourier transform of the operator that adds a unit of
angular
momentum at site $j$.  Using Eq.~(\ref{twenty})
\begin{equation}
[Q_j,b_{\vec{q}}]=\sqrt{U/N}e^{i\vec{q}\cdot {\vec R}_j}e^{i\theta_j}
\label{twentyone}
\end{equation}
it follows that $f_{\vec{q}}=\langle\Psi_0\vert b_{-\vec{q}}Hb_{\vec{q}}
\vert\Psi_0\rangle=U$.  The static structure factor is given by the spin
susceptibility of the classical XY model at wave vector $\vec{q}$
\begin{equation}
s(\vec{q})={1\over
N}\sum_{ij} e^{i\vec{q}\cdot ({\vec R}_i-{\vec R}_j)}\langle\Psi_0\vert
e^{i(\theta_i-\theta_j)}\vert\Psi_0\rangle \label{twentythree}
\end{equation}
which we know from Kosterlitz-Thouless theory:
\begin{equation}
\langle
e^{i\theta_i}e^{-i\theta_j}\rangle\sim\begin{array}{ll} \vert {\vec
R}_i-{\vec
R}_j\vert^{-\eta}\qquad & T<T_c\\ e^{-\vert {\vec R}_i-{\vec
R}_j\vert/\xi}\qquad & T>T_c \end{array}
\label{twentyfour}
\end{equation}
where $T_c$ is the critical temperature, or critical coupling $(U/J)_c$ in
our
case, with $\eta$ ranging from 0 to 1/4 as temperature varies from 0 to
$T_c$.
In the spin-wave approximation $\eta=\frac{U}{4\pi J}$ for our model.
Substituting Eq.~(\ref{twentyfour}) in Eq.~(\ref{twentythree}), and
changing
summations to integrals for an infinitely large system with $U/J$ below the

critical point
\begin{equation}
s(\vec{q})\sim\frac{1}{2\pi}\int_0^\infty r^{-\eta
+1} J_0(qr)dr\sim q^{-2+\eta } ,
\label{twentysix}
\end{equation}
where $r=\vert {\vec R}_i-{\vec R}_j\vert$ and since $f(\vec{q})=U$
\begin{equation} \Delta (\vec{q})\sim\frac{U}{s(\vec{q})}\sim Uq^{2-\eta} .
\label{twentyseven}
\end{equation}
For $U/J$ above the critical point and $q\xi\ll 1$
\begin{equation}
s(\vec{q})=\frac{1}{2\pi}\int_0^\infty drre^{-r/\xi}J_0(qr)\sim\xi^2
\label{twentyeight}
\end{equation}
and hence the quantum system has an excitation gap (within the single mode
approximation)
\begin{equation}
\Delta (\vec{q}=\vec{0})\sim U\xi^{-2} .
\label{thirty}
\end{equation}
Using the Kosterlitz-Thouless theory\cite{kosterlitz} prediction for the
correlation length we have
\begin{equation}
\Delta(0)\sim\exp{\left\lbrace -2b\Bigl[ {J\over U}-\biggl(
\frac{J}{U}\biggr)_c\Bigr]^{-1/2}\right\rbrace }
\label{thirtytwo}
\end{equation}
where $b$ is a positive constant.

Figs.~(\ref{fig1}-\ref{fig2}) illustrate the basic features of the
excitation
energy.   In the thermodynamic limit the system is gapless below the
critical
point and has a  gap which rises from zero above the critical point with
the
essential  singularity characteristic of the KT transition.  In the
ordinary
quantum XY model, the Bogoliubov process mixes the single-particle and
density
excitations to produce a linearly dispersing collective Goldstone mode in
contrast to the $\omega\sim q^2$ dispersion of free bosons.  We see here
from
Eq.~(\ref{twentyseven}) a curious contrast to the generic behavior.  The
collective mode dispersion $\omega\sim q^{2-\eta}$ gradually stiffens with
increasing $U/J$ but never becomes linear since $\eta\leq 1/4$ below the
transition.

We note that right at the critical point the $q=0$ single-particle energy
only
vanishes in the thermodynamic limit $L\rightarrow\infty$ since the
classical XY
model susceptibility obeys $s(\vec{q})\sim\int d^2r\; r^{-1/4}\sim
L^{7/4}$.
Thus we expect $L^{7/4}\Delta (\vec{0})$ to be scale-invariant (i.e.
independent
of $L$) at the critical point, provided $L$ is large enough.  Using data
from
lattices with $8\leq L\leq 24$ we found the scale-invariant point to be
$U/J\sim
0.905$.  On general renormalization group grounds we expect logarithmic
corrections to scaling for small $L$ in the 2D XY model.\cite{olsson}  We
include these corrections in Fig.~\ref{fig3} where we plot $L^{7/4}\Delta
(\vec{0})/[1 + 1/(2\ln{L} + 4.5)]$ vs $U/J$.  We again find the critical
value
$U/J\sim 0.905$ but the scaling now works well all the way down to $L=4$.
Our
value for the critical coupling is close to, but somewhat above, the value
of
$U/J\sim 0.895$ found by Olsson and Minnhagen\cite{olsson} using the
scaling of
the superfluid density.

We turn now to a study of the collective density mode excited state by
taking
$\rho_{\vec{q}}$ to be the Fourier transform of the number density
\begin{equation}
\rho_{\vec{q}}=\sum_i e^{i\vec{q}\cdot \vec{R}_i}(-i\partial_{\theta_i})
\label{twentytwo}
\end{equation}
so that
\begin{equation}
\Psi_{\vec{q}}=\frac{iJ}{U}\sum_{i\delta} e^{i\vec{q}\cdot {\vec R}_i}\;
\lbrace\sin{(\theta_i-\theta_{i+\delta})} + \sin{(\theta_i -
\theta_{i-\delta})}
\rbrace\Psi_0 .
\label{thirtythree}
\end{equation}
The static structure factor is
\begin{eqnarray}
s(\vec{q})&=&\frac{J^2}{U^2}\sum_{ij\delta\delta '} e^{i\vec{q}\cdot
({\vec R}_i-{\vec R}_j)}\;\langle\Psi_0\vert\;
[\sin{(\theta_i-\theta_{i+\delta}
)} +\sin{(\theta_i-\theta_{i - \delta})}]\nonumber\\
& & [\sin{(\theta_j - \theta_{j+\delta '})} +
\sin{(\theta_j-\theta_{j - \delta'})}]\vert\Psi_0\rangle .
\label{thirtyfour}
\end{eqnarray}
Following the steps in the calculation of oscillator strength for the
single-particle model, we find $f(\vec{q})$ for the density wave state to
be
\begin{eqnarray}
f(\vec{q})&=&\langle\Psi_0\vert {J^2\over U}\sum_{j\delta\delta
'}\lbrace(1-e^{i\vec{q}\cdot {\vec\delta}})(1-e^{-i{\vec q}\cdot
{\vec\delta}'})\cos{(\theta_j-\theta_{j+\delta})}\cos{(\theta_j-\theta_{j+\d
elta
'})}\nonumber\\
& &+(1-e^{-i\vec{q}\cdot {\vec\delta}})(1-e^{-i\vec{q}\cdot
{\vec\delta}'})\cos{(\theta_j-\theta_{j-\delta})}\cos{(\theta_j-\theta_{j+\d
elta
'})}\nonumber\\
& &+ (1-e^{i\vec{q}\cdot {\vec\delta}})(1-e^{i\vec{q}\cdot
{\vec\delta}'})\cos{(\theta_j-\theta_{j+\delta})}\cos{(\theta_j-\theta_{j-\d
elta
'})}\nonumber\\
& &+ (1-e^{-i\vec{q}\cdot {\vec\delta}})(1-e^{i\vec{q}\cdot
{\vec\delta}'})\cos{(\theta_j-\theta_{j-\delta})}\cos{(\theta_j-\theta_{j-\d
elta
'})}\rbrace\vert\Psi_0\rangle .\label{thirtyfive}
\end{eqnarray}
We performed Monte Carlo simulations of the 2D XY model to find the excited

state energy $\epsilon_{\vec{q}}\leq f(\vec{q})/s(\vec{q})$ for systems of
finite size.  Though one expects some physical connection between
single-particle and density-mode approximations, our results for the two
models
are quite different.  We found out that unlike single-particle excitation,
the
density wave is nearly dispersionless.  Consequently, the results do not
change
significantly with the system size in the latter case. The excited state
energy
$\epsilon_{\vec{q}}$ vs. $\vec q$ is shown in Fig.~(\ref{fig4}), with
$\vec{q}$
in (1,0) direction.  Fig.(\ref{fig5}) is a plot of $\epsilon_{\vec{q}}$ vs.
$U/J$ for the smallest  non-zero allowed vector on a $32\times 32$ lattice,
i.e.,  $\biggl( \frac{2\pi}{32}, 0\biggr)$. We notice that the excited
state
energy for this model increases gradually with coupling $U/J$, contrary to
the
single-particle case, where the energy is close to zero below the critical
point, and then abruptly increases.

We conclude with some comments on additional curious features of this
model. The
model is readily extended to include exact solutions for arbitrary random
bond
strengths $J_{i,\delta}$ and frustration vector potential $A_{i,\delta}$.
One
might imagine that since the ground state energy is identically zero
independent of the disorder realization, one could compute
ensemble-averaged
correlation functions without having to invoke the replica
trick.\cite{goldenfeld}  This is not possible however since the {\em norm}
of
the  ground wave function
 \begin{equation}
\Phi (\theta_1,\cdots ,\theta_N)=e^{\sum_{j\delta}
\frac{J_{j,\delta}}{U}\cos{(\theta_j - \theta_{j+\delta} + A_{j\delta})}}
\label{thirtysix}
\end{equation}
{\em does} depend on the disorder.  A second consequence of the ground
state
energy being zero for all  $A_{j,\delta}$ is that {\em the superfluid
density is
identically zero} at $T=0$  {\em even though the system exhibits
(algebraic)
ODLRO below the critical value of} $U/J$.

\acknowledgements
We are grateful for useful discussions with D.P. Arovas, M. Wallin, and
K. Mullen.  This work was supported by NSF DMR-9113911.

\begin{figure}
\caption{Excitation energy $\Delta_{\vec{q}}$ as a function of $\vec{q}$
for
$U/J=0.8$ and 0.9.  The ($\times$) are for $\vec q$ in the (1,0) direction
and
the ($\diamondsuit$) are for $\vec q$ in the (1,1) direction.  For small
$\vec
q$, the single particle model is isotropic.}
\label{fig1}
\end{figure}

\begin{figure}
\caption{Single particle excitation energy vs. coupling at $\vec{q}=0$, for
different system sizes. Notice that $\Delta_{\vec{q}}\rightarrow 0$ as the
system size increases for $U/J$ below the critical point.}
\label{fig2}
\end{figure}

\begin{figure}
\caption{Estimation of critical coupling for the single-particle model.
$L^{7/4}\Delta (\vec{0})$ is independent of system size $L$ {\em at\/} the
critical point according to the Kosterlitz-Thouless theory. The logarithmic
factor is a correction to scaling (see text).  The parameter $C\sim 4.5$
was
adjusted to obtain the best scaling.}
\label{fig3}
\end{figure}

\begin{figure}
\caption{Excitation energy $\epsilon_{\vec{q}}$ vs. $\vec q$ for
single-mode
density-wave approximation. $\vec q$ is in the (1,0)  direction. The value
of
the coupling $U/J$ is shown next to each curve.}
\label{fig4}
\end{figure}

\begin{figure}
\caption{The density wave excitation energy $\epsilon_{\vec{q}}$ increases
gradually as a function of $U/J$ in the single-mode approximation. The
results
are for  $\vec{q}=\biggl(\frac{2\pi}{32},0\biggr)$, the smallest non-zero
allowed wave  vector on a $32\times 32$ lattice.}
\label{fig5}
\end{figure}


\begin{references}
\bibitem{referencea} To avoid douple counting $\delta$ is summed on half
the
near-neighbors.  For example in two-dimensions $\delta$ is summed on
neighbors
above and to the right of $j$.
\end{references}
\end{document}